\documentclass[]{article}
\usepackage[a4paper]{geometry}
\usepackage{url}
\usepackage{array}

\usepackage{mathtools}
\usepackage{hyperref}
\usepackage{amsmath}
\usepackage{subcaption}
\usepackage{authblk}
\usepackage{xspace}
\usepackage{comment}
\usepackage{lineno}

\newcommand{\p}{\ensuremath{p}} 

\newcommand{\tr}{\mathrm{tr}}
\newcommand{\pI}{\ensuremath{\mathrm{I}}\xspace}
\newcommand{\pX}{\ensuremath{\mathrm{X}}\xspace}
\newcommand{\pY}{\ensuremath{\mathrm{Y}}\xspace}
\newcommand{\pZ}{\ensuremath{\mathrm{Z}}\xspace}
\newcommand{\Rx}{\ensuremath{\mathrm{R}_x}}
\newcommand{\Ry}{\ensuremath{\mathrm{R}_y}}
\newcommand{\Rz}{\ensuremath{\mathrm{R}_z}}
\newcommand{\Dtrace}{\ensuremath{D_{\mathrm{trace}}}}
\newcommand{\F}{\ensuremath{F}} 
\newcommand{\code}[1]{\texttt{#1}}
\newcommand{\nq}{k} 
\newcommand{\niq}{l} 
\newcommand{\rsection}[1]{{\bfseries #1}}
\newcommand{\shots}{\ensuremath{N}}
\newcommand{\bO}{\ensuremath{O}} 

\newcommand{\equationref}[1]{Equation~\ref{#1}}
\newcommand{\tableref}[1]{Table~\ref{#1}}
\newcommand{\figureref}[1]{Figure~\ref{#1}}
\newcommand{\Figureref}[1]{Figure~\ref{#1}}
\newcommand{\appendixref}[1]{Appendix~\ref{#1}}

\usepackage{framed}
\usepackage{xcolor}
\definecolor{purple}{rgb}{0.5,0,1}

\excludecomment{bcomment}

\title{Z-basis measurements using mixed parity and direct readout}
\author[1]{{Pieter Thijs} {Eendebak}}
\author[1]{\"Onder {G\"ul}}
\affil[1]{QuTech and Netherlands Organization for Applied Scientific Research (TNO), Stieltjesweg 1, 2628 CK Delft, Netherlands}
\begin{document}
	\maketitle
	\begin{abstract}
		Many architectures for quantum information processing rely on qubits dedicated for the readout of a larger quantum register. These ancilla readout qubits present a physical overhead not contributing to the computational resource. A common implementation in spin qubit architectures is the readout schemes based on Pauli exclusion of charges confined in a double quantum dot, with one dot serving as the ancilla qubit. Here, using a three-qubit spin register and a Pauli exclusion-based readout, we present z-basis measurements of the entire register constructed with tomography, eliminating the physical overhead. We validate our approach with simulations which provide insight into potential sources of errors in the reconstruction. We also demonstrate our reconstruction by performing quantum state tomography on a GHZ state of a spin-qubit based device.
		
		\begin{bcomment}
			Note: text in blue is for the authors and reviewers. This will be removed or rewritten in the final version of the paper.
		\end{bcomment}
	\end{abstract}
	
	\section{Introduction}

	Recent demonstrations of spin-qubit devices~\cite{Philips2022, Hendrickx_2021, Weinstein2023, Huang2024, } rely on Pauli exclusion-based spin-to-charge conversion\cite{Ono2002} due to its inherent advantages in scalability.
	For timescales relevant for readout, this so-called Pauli spin blockade (PSB) allows for measuring the spin parity of a double quantum dot~\cite{Seedhouse_2021} which operate as two Loss-DiVincenzo qubits~\cite{Loss1998}. A disadvantage of this approach is the typical sacrifice of one of the dots as an ancilla dot for readout which decreases the quantum volume~\cite{Cross2019a}. Here we mitigate this disadvantage with computational overhead and realize a z-basis measurement of the complete register.
	
	In principle, using only parity measurements on a pair of qubits allows for the reconstruction of the full state of the entire register. A recent study~\cite{Seedhouse_2021} has demonstrated the construction of the density matrix of a two-qubit system using parity measurements. However, for larger systems with $k$ qubits, the degree of freedom in the density matrix grows as $2^{2k} - 1$ and many circuits are needed to reconstruct the full density matrix. In contrast, a z-basis measurement on a $k$ qubit system contains only $2^k -1$ degrees of freedom and will require significantly less circuits to perform a reconstruction.
	
	We focus on a common configuration of a linear array of spin qubits~\cite{Philips2022}; however, our technique is applicable to other configurations as well, for example to arrays of $n \times n$ ~\cite{Hendrickx_2021}. In a linear array, the readout is typically performed by a parity measurement on the outer qubit pair where the outermost qubit is used as an ancilla~\cite{Seedhouse_2021}. After the parity measurement, the remaining inner qubits are readout by a state transfer using CROT gates followed by another the parity readout on the outer qubit pair. This readout scheme does not prevent the use of the outer qubits as computational qubits. We direct our attention to the left-hand side of the device, that is qubit 0, 1 and 2. The readout of the right-hand side can be performed similarly, in parallel or sequentially.

	For a device with $\nq$ qubits, a single circuit with parity measurement on the outer qubits and direct measurements on the other yields $2^{\nq-1} -1$ bits of information per shot. The minimum number of circuits required to perform the reconstruction can be determined considering the bits of information per shot  and the degrees of freedom. For the z-basis reconstruction, the minimum number of circuits needed is 3; for the density matrix reconstruction this is $\lceil{(2^{2\nq} - 1)}/{(2^{\nq-1} -1)}\rceil \geq 2^{\nq+1}$ (\appendixref{appendix:dof}).

	\section{Methods}

	In the introduction we have already seen we need at least 3 circuits to perform tomographic reconstruction of the z-basis measurements. We now proceed with choosing the three circuits to perform the tomographic reconstruction of the z-basis measurements. Our goal is to reconstruct the values $|a_j|^2$ for a state $\sum_j a_j |j\rangle$ using tomography similar to quantum state tomography~\cite{Smolin2012}. A key difference is that we reconstruct the values $|a_j|^2$ and not the full density matrix or coefficients $a_j$.

\begin{figure}
		\centering
		\includegraphics[width=0.9\linewidth]{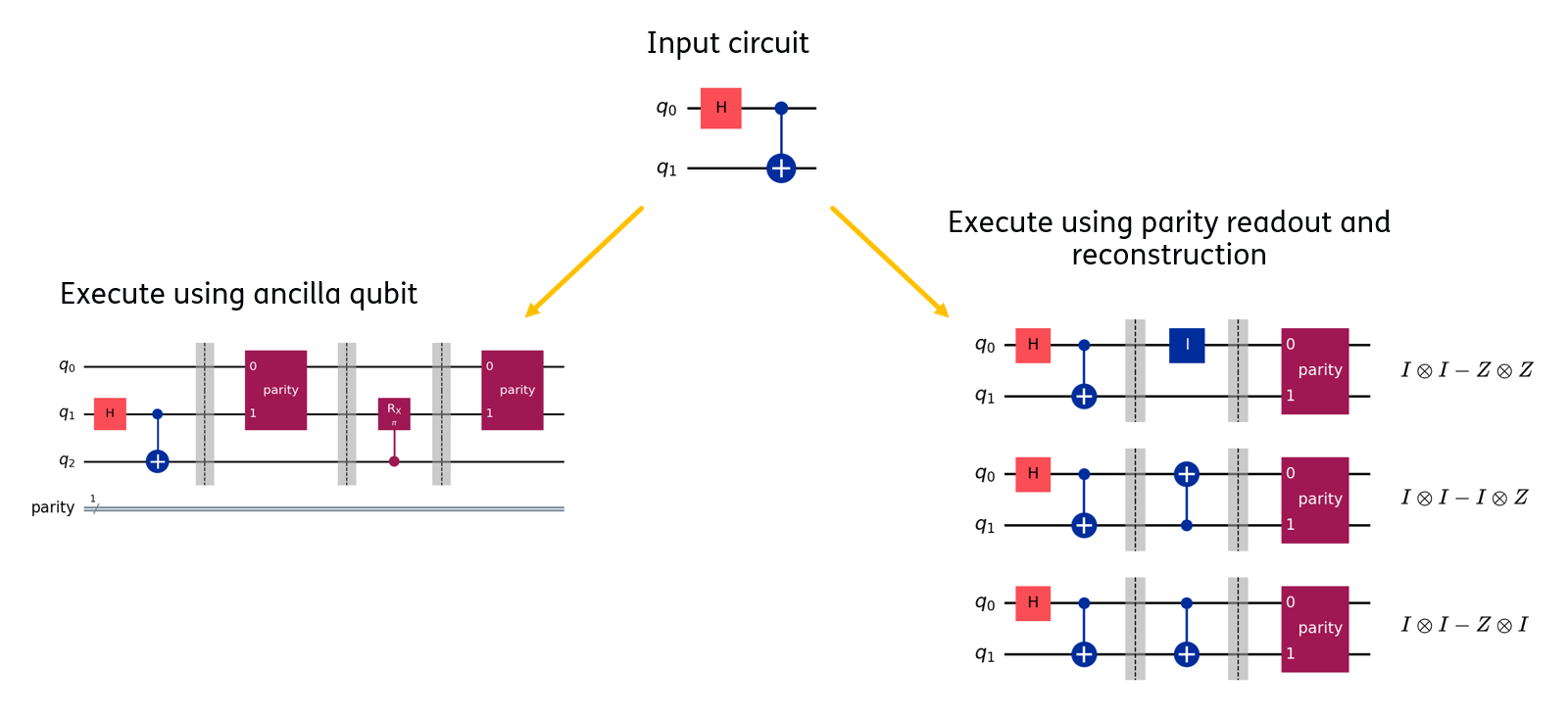}
		\caption{Measuring a Bell state. a) Using an ancilla qubit for readout b) Using reconstruction using parity readout}
		\label{fig:2qubittomographycircuits}
	\end{figure}

	Note that every observable can be expressed as a sum of Pauli terms (see~\cite{enwiki:pauli_matrices}), e.g.~as a sum of terms from $\{\pI,\pX,\pY,\pZ\}^{\otimes k}$.
	As described in \appendixref{appendix:reconstruction circuits} we can restrict the choice of our observables to  a linear combination of the terms  $\pI \otimes \pI$, $\pI \otimes \pZ$,  $\pZ \otimes \pI$,  and $\pZ \otimes \pZ$.
	From a reconstruction point of view, any set of three linear independent observables is equally suitable. We therefore choose the a set of observables for the outer qubits which can be implemented with only few gates in a broad range of hardware systems and are symmetric with respect to the qubit pair involved in the parity readout:
		\begin{itemize}
			\item $\bO_1 = \pI \otimes \pI - \pZ \otimes \pZ$: Parity measurement (no additional gates)
			\item $\bO_2 = \pI \otimes \pI - \pI \otimes \pZ$: Measurement on first qubit (requires a single CNOT)
			\item $\bO_3 = \pI \otimes \pI - \pZ \otimes \pI$: Measurement on second qubit (requires a single CNOT)
		\end{itemize}	
	These observables correspond to the circuits shown in \figureref{fig:2qubittomographycircuits}.
		
\begin{figure}
		\centering
		\includegraphics[width=0.87\linewidth]{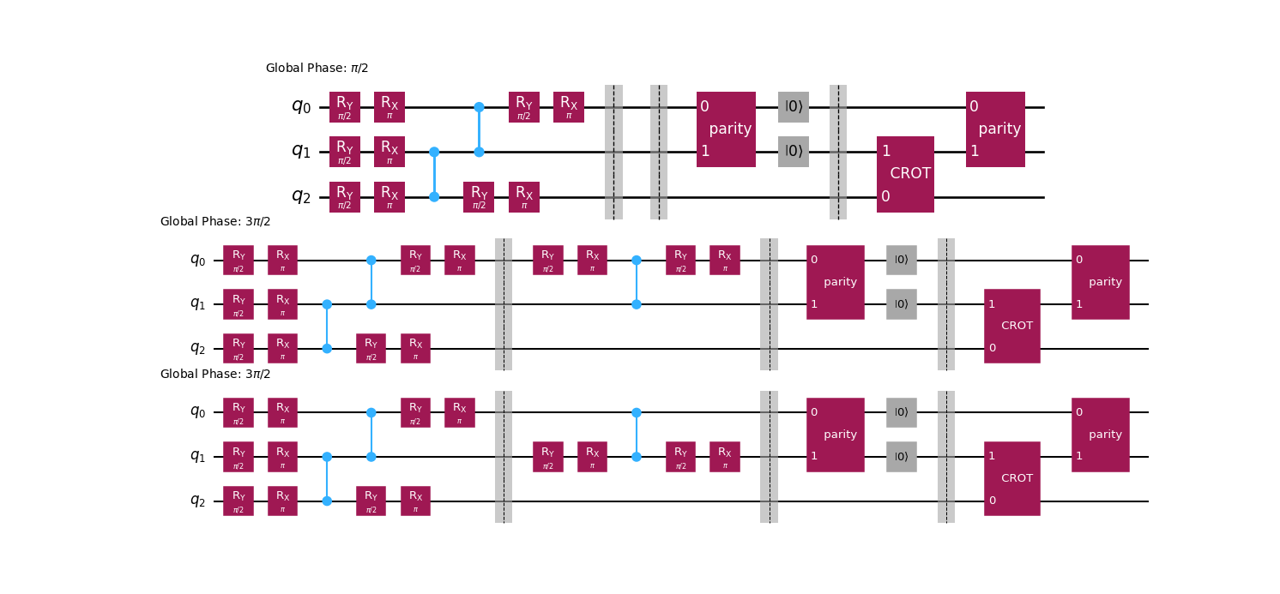}
		\caption{Tomography circuits for a 3-qubit GHZ state. The readout of the inner qubit is via a controlled rotation.}
		\label{fig:3qubitdevicetomographycircuits}
	\end{figure}

	For a linear array with parity readout on the outer qubits and direct readout on the inner qubits, we
	can take the terms $(\pI \pm \pZ)/2$ for each of the inner qubits. This leads to the full
	set of $3 \cdot 2^{k-1}$ observables that can be measured with only 3 circuits:
 $\bO_1 \otimes ((\pI \pm \pZ)/2)^{\otimes \niq}$, 
 $(I \otimes I - \bO_1) \otimes ((\pI \pm \pZ)/2)^{\otimes \niq}$,
 $\bO_2 \otimes ((\pI \pm \pZ)/2)^{\otimes \niq}$, 
 $(I \otimes I - \bO_1) \otimes ((\pI \pm \pZ)/2)^{\otimes \niq}$,
 $\bO_3 \otimes ((\pI \pm \pZ)/2)^{\otimes \niq}$, 
 $(I \otimes I - \bO_1) \otimes ((\pI \pm \pZ)/2)^{\otimes \niq}$,
   (here $\niq=\nq-2$ is the number of qubits with direct readout).
	 \figureref{fig:3qubitdevicetomographycircuits} shows the circuits expanded for a three-qubit spin device. 
	
	\begin{figure}
		\centering
		\includegraphics[width=0.87\linewidth]{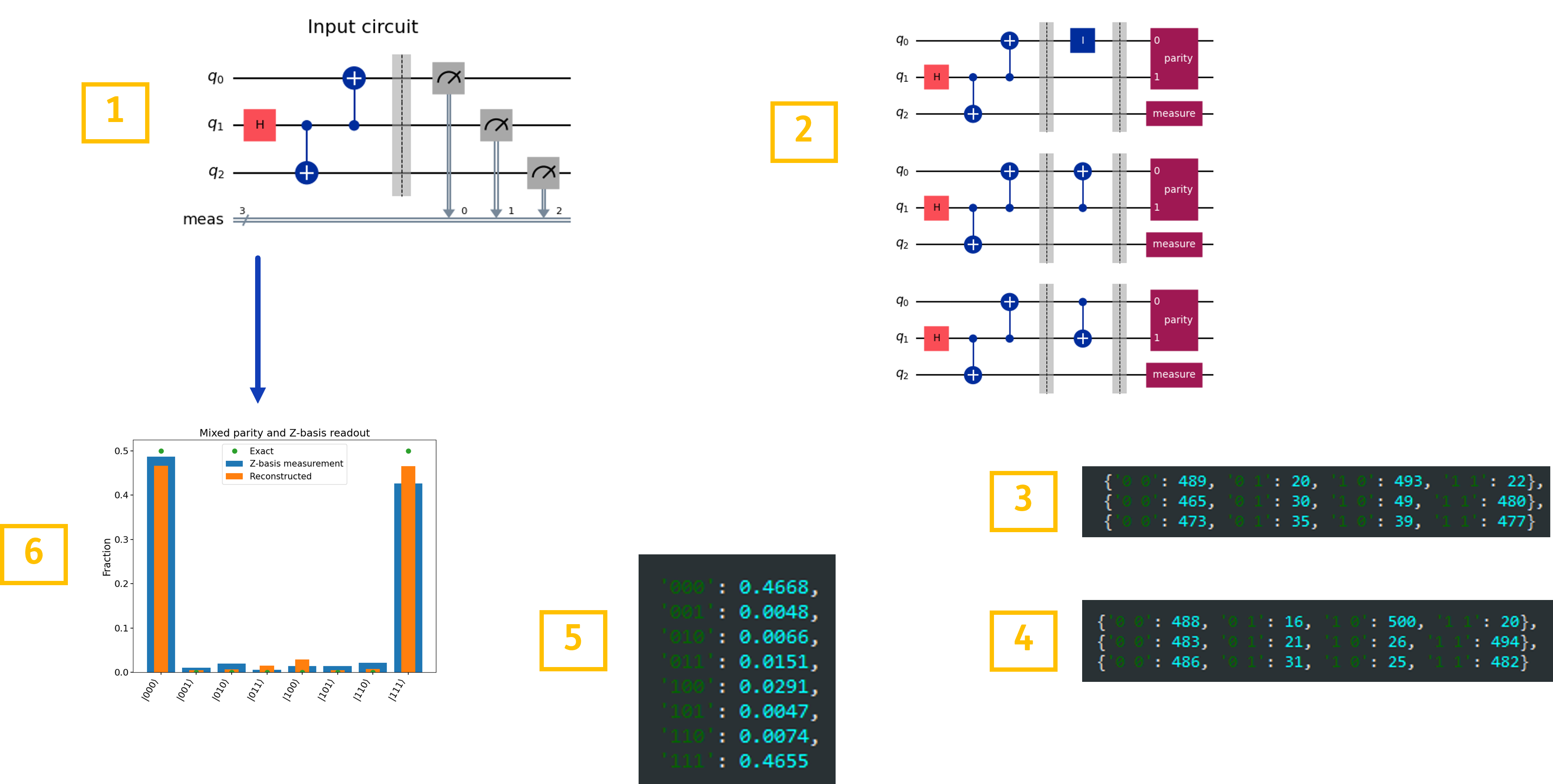}
		\caption[Tomography steps]{Different steps in the reconstruction. The user input is a normal circuit diagram (1). The diagram is converted to three tomography circuits (2), executed on the backend to obtain counts (3). In post-processing readout error mitigation is applied (4) and the expectation values for the observables calculated. Using the expectation values and the inversion matrix $M$ the reconstructed counts are calculated (5) and returned to the user (6).}
		\label{fig:paritytomographycycle}
	\end{figure}

	The linear inversion matrix $M$ maps the coefficients $|a_j|^2$ to the expectation values of the observables and has shape $3 \cdot 2^{\nq-1} \times 2^\nq$. The matrix for $\nq=2$ is given by
	\begin{align}
		M &= \begin{pmatrix}
			1 & 0 & 0 & 1  \\
			0 & 1 & 1 & 0  \\
			1 & 0 & 1 & 0  \\
			0 & 1 & 0 & 1  \\
			1 & 1 & 0 & 0  \\
			0 & 0 & 1 & 1  
		\end{pmatrix},
	\end{align}
	and the matrix for $k=3$ is given in~\equationref{equation:M3}.

    
	Because the matrix has rank equal to the number of columns, we can apply least-squares to reconstruct the z-basis from the expectation values. Further, our linear reconstruction results in correct estimates of the z-basis measurements also for a system in a mixture state (e.g.~a 50/50 mixture of $|00\rangle$ and  $|11\rangle$). However, the least-squares estimate of our solution can lead to non-physical estimates (probabilities outside the range $[0, 1]$ can be generated).
	To mitigate this, we perform a second optimization step, a gradient descent optimization~\cite{ruder2017} with initial point the least-squares solution and the following heuristic cost function $J$:
	\begin{align}
		R &= 	(Mp - E_i)^2 \hfill \tag{residual} \\ 
		J(\p) &=  \sum_i R_i / \sigma^2_i + \alpha P(\p)
	\end{align} 
	Here, $E$ is the vector of measured values for the observables, $\sigma_i^2 = E_i (1-E_i)/\shots$ is the variance of the $i$th component of the measured observables, $\alpha$ the weight of the penalty term $P(p)$ and $R$ the residual weighted by the expected variance of the measurement $E$. The penalty term is zero for probabilities in the region $[0, 1]$ and greatly increases when outside. To avoid numerical instabilities near probabilities 0 and 1, we take a regularized version of the variance $\sigma_i^2$.  For details on the penalty term and regularization see appendix~\ref{appendix:regularization}.
	
	To determine the quality of our reconstruction method we use simulations and compare the reconstructed probabilities with the 
	exact probabilities using the total variation distance (or trace distance) as our fidelity metric. For two probability distributions $a$, $b$ the trace distance is defined as 
	\begin{align}
		\Dtrace(a, b) &= \sum_i |a_i - b_i| / 2  .
	\end{align}
	For details and the relation to the state fidelity see \appendixref{appendix:metrics}.

	\section{Results}
	
	We have analyzed the quality of our reconstruction method using simulations in Qiskit~\cite{qiskit2024} and by running it experimentally on a spin qubit device (see~\figureref{fig:6d2sdevicelayoutannotated}, equivalent to that in Ref.~\cite{Philips2022}). We first use simulations for the following methods: GHZ state creation, quantum state estimation, assignment matrix error and gate set tomography. These methods evaluate the performance of the readout using the reconstruction in comparison with alternative readout methods. Since our reconstruction method does not change the initialization or manipulation, we focus on the readout only. Our simulations are either noiseless or with noise representative for a spin-based quantum processor (details in Appendix~\appendixref{appendix:simulation details}). Next, we compare our reconstruction method with readout using an ancilla qubit. Finally, we create a 3-qubit GHZ state on a spin qubit based device.
	
\begin{figure}
		\centering
		\includegraphics[width=0.87\linewidth]{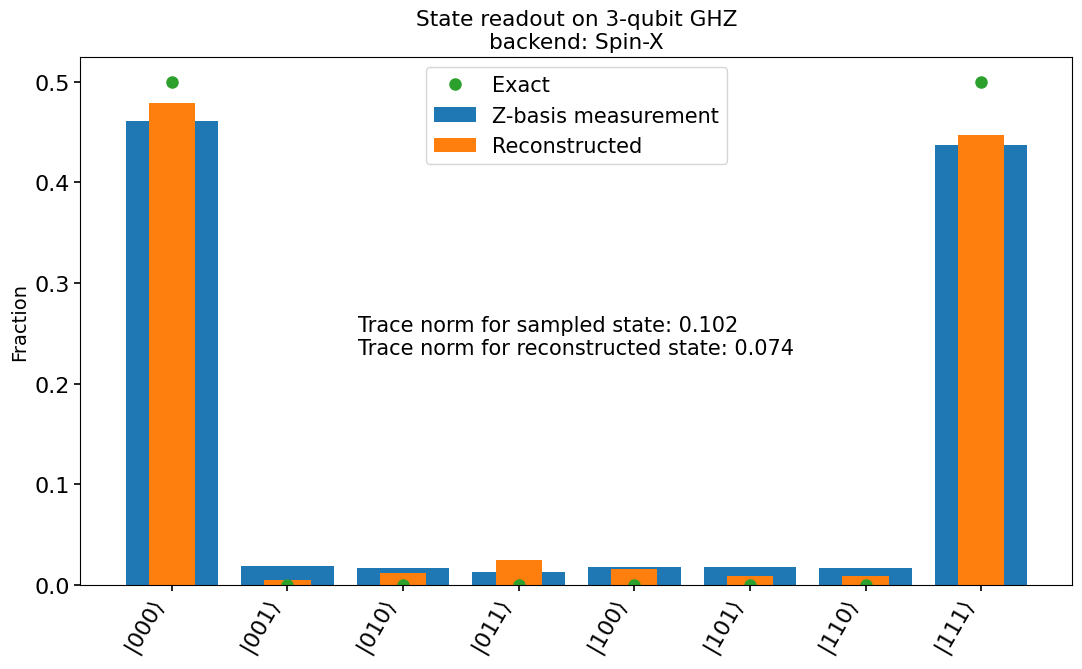}
		\caption{Reconstruction of a 3-qubit GHZ state.}
		\label{fig:reconstructionresult}
	\end{figure}

	\rsection{GHZ state creation}  
	The classical example of quantum entanglement is the Bell state whose multi-qubit generalization is the GHZ state. We therefore first apply our reconstruction method to a 3-qubit GHZ state. \figureref{fig:reconstructionresult} shows the results of a direct measurement and reconstruction. We note the trace norm of the reconstruction being lower than the direct measurement. This can be explained by considering the variance in the different circuit measurements.
	\begin{figure}
		\centering
		\includegraphics[width=0.85\linewidth]{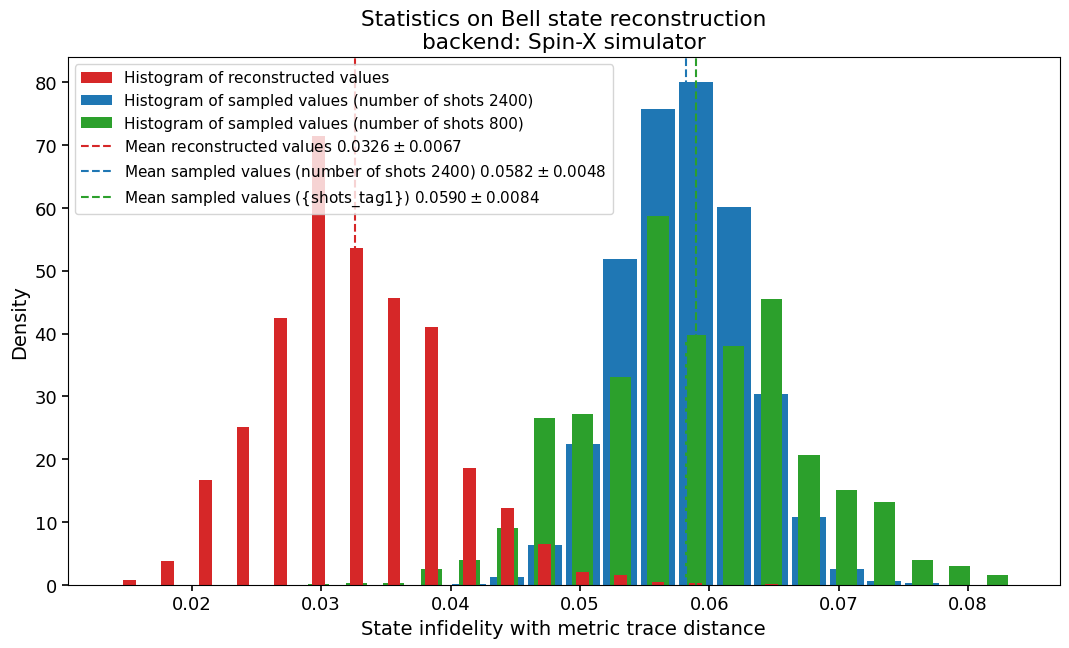}
		\caption[Reconstruction fidelity on Bell state]{Histogram of reconstruction fidelities on a Bell state. For a large number of simulations we
		calculated the trace distance from the exact probabilities to the reconstructed probabilities (red).
		We also calculated a histogram of the trace distances
		between sampled probabilities for 800 and 2400 shots (green and blue).  }
		\label{fig:tomographystatisticsvigobell}
	\end{figure}
	The variation in a state measurement (or reconstruction) is determined by the state and the number of shots.
	When measuring a 3-qubit GHZ state $|000\rangle + |111\rangle$ the variance in the bitstring fractions \code{000} and \code{111} is $.25/\shots$. This variance results in a non-zero trace norm for direct measurements. For the  reconstruction using parity readout method, the first circuit measures the parity of the GHZ state which is even, hence this measurement has no variation (for an ideal device).
	The low variation in this first circuit leads to a better fidelity for the  reconstruction than for a normal measurement. When performing the measurements on a large number of identically prepared GHZ states, we see that indeed the reconstruction method performs better on average (\figureref{fig:tomographystatisticsvigobell}). To prevent such state-dependent bias, we analyze the performance of the reconstruction on random unitary states instead of a single state.
	
\begin{figure}
		\centering
		\begin{subfigure}{0.85\linewidth}\centering
		\includegraphics[width=0.85\linewidth]{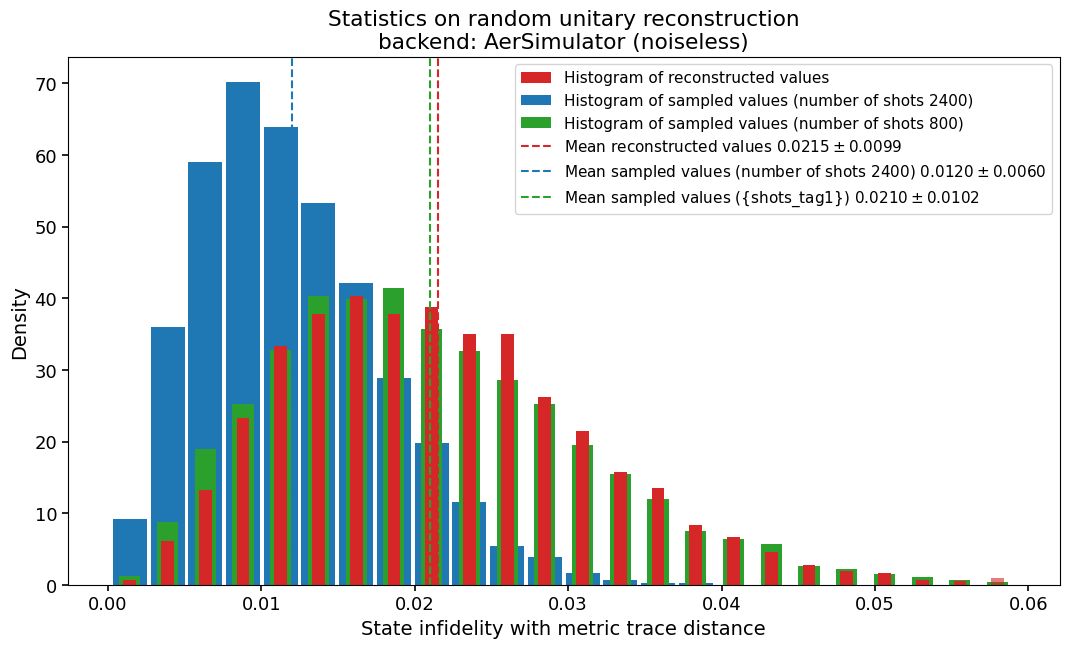} \centering
		\caption{Qiskit Aer (noiseless)}
		\label{fig:tomographystatistics_aer}
		\end{subfigure} 
		\begin{subfigure}{0.85\linewidth}\centering 
		\includegraphics[width=0.85\linewidth]{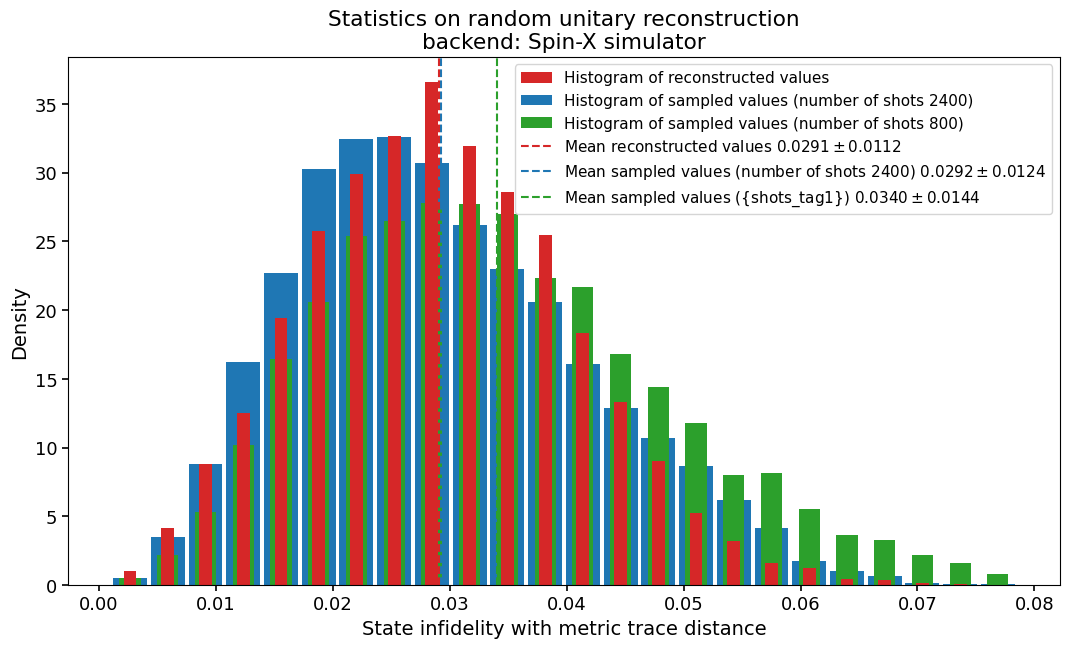}
		\caption{Spin-X backend simulator}
		\label{fig:tomographystatistics_spinx}
		\end{subfigure}
	\caption{Reconstruction results on random unitary gates. All histograms have been obtained with 8000 runs. The reconstruction results have been obtained with 800 shots per circuit.}
	\end{figure}

	\rsection{State estimation results} For the reconstruction method we run our circuits with 800 shots. To compare with the direct readout method we simulate the direct method both with 800 shots as well or with 2400 shots (the total number of shots being equal for both methods in the latter).
	\begin{bcomment}
		For real devices a better measure would perhaps be the total execution time. However, this depends on compilation overhead, the duration of the gates, measure statements and various other factors. For real devices we expect the equivalent execution time ratio to be between 2 and 3.
	\end{bcomment}
	Figures~\ref{fig:tomographystatistics_aer} and \ref{fig:tomographystatistics_spinx} show the results of a simulation. While for the noiseless simulation the reconstruction method performs worse than direct measurement with 2400 shots and comparable to direct measurement with 800 shots, the methods perform comparable for the simulation with noise (see appendix~\ref{appendix:simulation details} for details). The reduction of the impact of the number of shots for the direct measurements is due to other noise sources becoming more important. We note that the quality of the readout depends on the simulation settings such as readout fidelity, which greatly affects both the direct and the reconstruction method, and the quality of the CX gate which only affects the reconstruction method. 
	
\begin{figure}
		\centering
		{\includegraphics[width=0.46\linewidth]{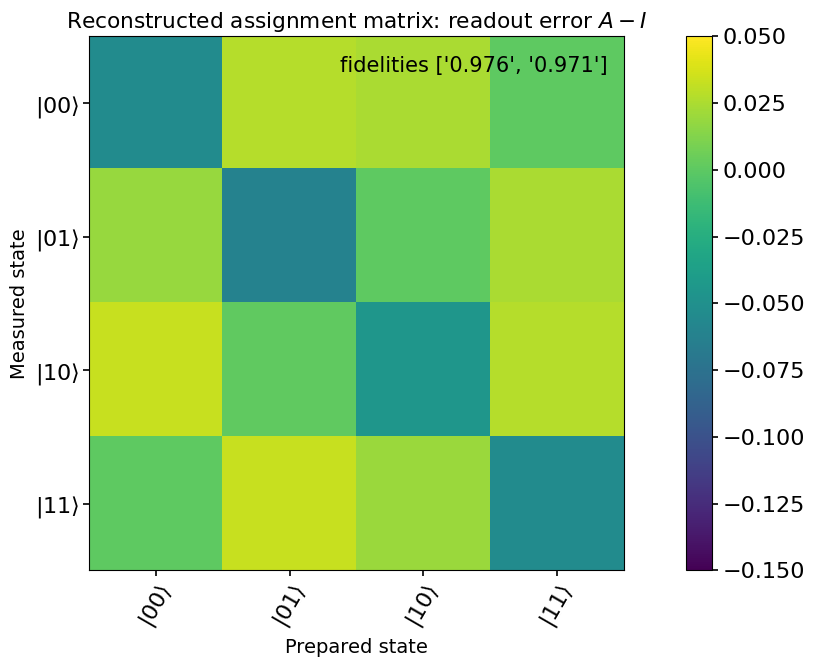}}
		\includegraphics[width=0.46\linewidth]{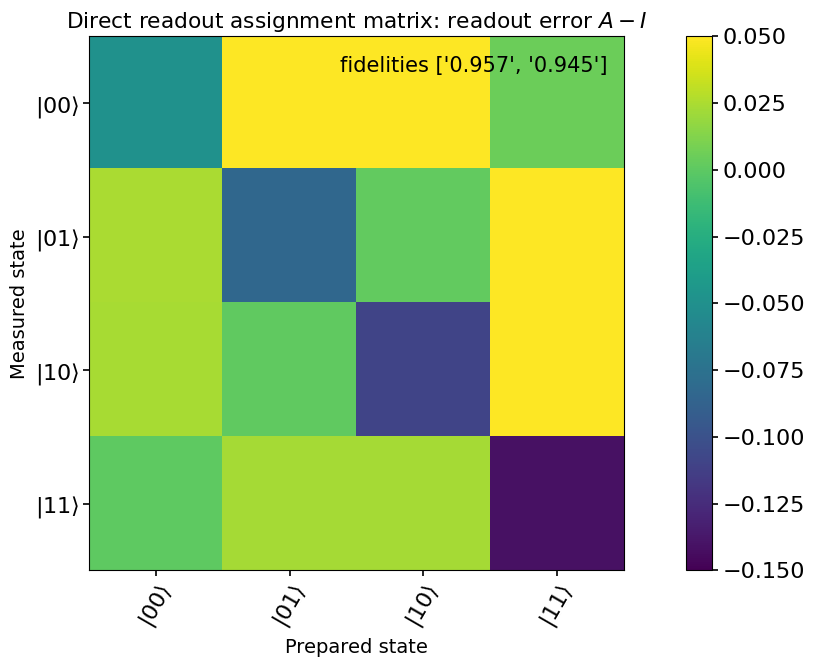}
		\caption{Readout assignment matrices for reconstruction using parity readout (left) and direct readout (right) for a simulation with ideal gates and readout noise. Displayed is the difference between the measured matrix and the ideal matrix (the identity matrix).}
		\label{fig:confusionmatrix}
	\end{figure}

	{\bfseries Readout error}  We have measured the readout assignment matrix~\cite{mitiqWhatTheory} using direct readout and reconstruction using parity measurements. For a noiseless simulation, both reconstruction methods are perfect (fidelity 1). For our Spin-X simulation, the fidelities of both qubits are slightly higher than those in the direct readout (\figureref{fig:confusionmatrix}). This is again a consequence of the state preparation for the readout error mitigation resulting in states with definite parity.
		
	\rsection{Gate set tomography} We have performed gate set tomography (GST) on the circuits measured with our parity reconstruction method. GST can reconstruct the gates (X90, Y90, CZ) and estimate the state preparation and measurement (SPAM) probability. Performing GST with our method results in the correct gate estimates, which validates our reconstruction method. The SPAM probability estimated from GST with our method is 0.957, which is close to the SPAM probability of 0.964 for direct readout, which further confirms our readout quality being comparable to that for direct readout (Appendix~\ref{appendix:gst}).

\begin{figure}
		\centering
		\includegraphics[width=0.9\linewidth]{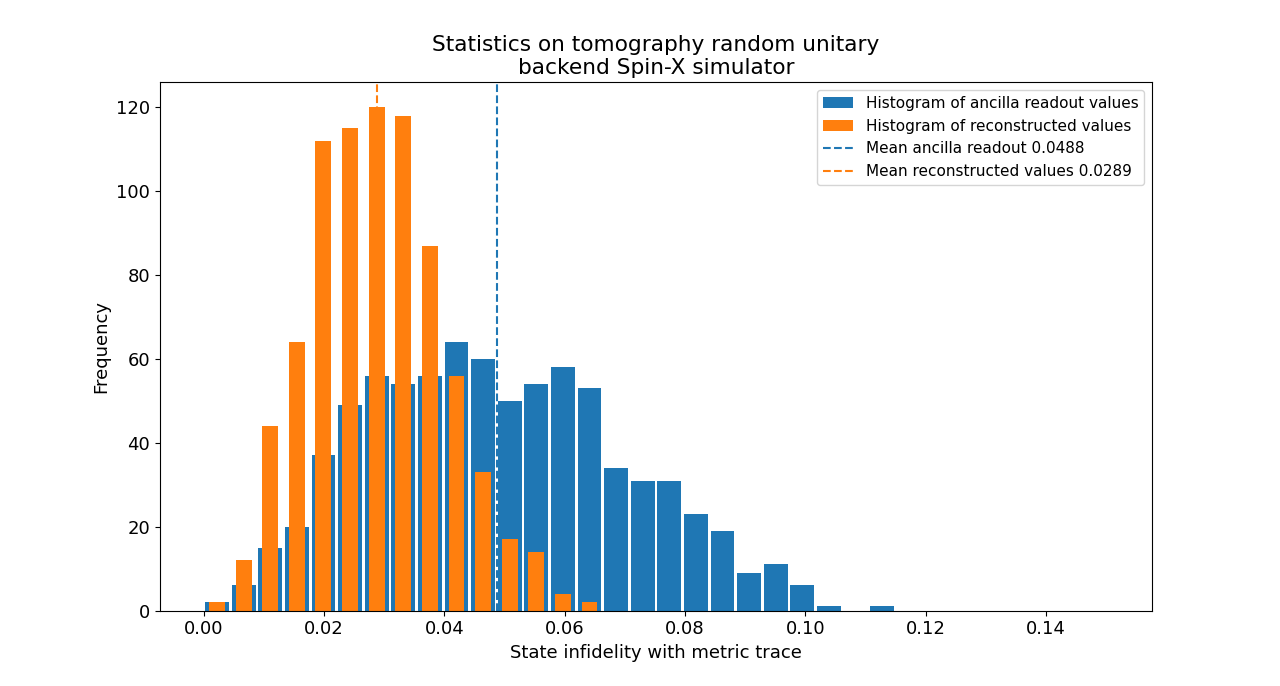}
		\caption{Reconstruction of random unitaries using ancilla based readout and reconstruction using parity readout.}
		\label{fig:ancilla readout}
	\end{figure}

	\rsection{Comparison of reconstruction using parity readout and readout using an ancilla qubit}
	For a physical quantum system implementation, the readout choice (e.g. ancilla based readout or reconstruction using parity readout, see \figureref{fig:2qubittomographycircuits}) is a trade-off between the resulting number of computational qubits, the resulting gate fidelities and the circuit execution time. This trade-off is heavily system dependent. Here we compare the two approaches which result in two computational qubits on a 3-qubit system. 
	\figureref{fig:ancilla readout} shows the comparison between the two methods. The reconstruction using parity readout performs better, suggesting its advantage over the ancilla based readout.

\begin{figure}
		\begin{subfigure}{0.45\linewidth}
			\centering
			\includegraphics[width=\linewidth]{3qubit\_ghz\_circuit}
			\caption{Input circuit}
			\label{fig:3qubitghzcircuit_a}
		\end{subfigure}\hspace{\fill}
		\begin{subfigure}{0.65\linewidth}
			\centering
			\includegraphics[width=\linewidth]{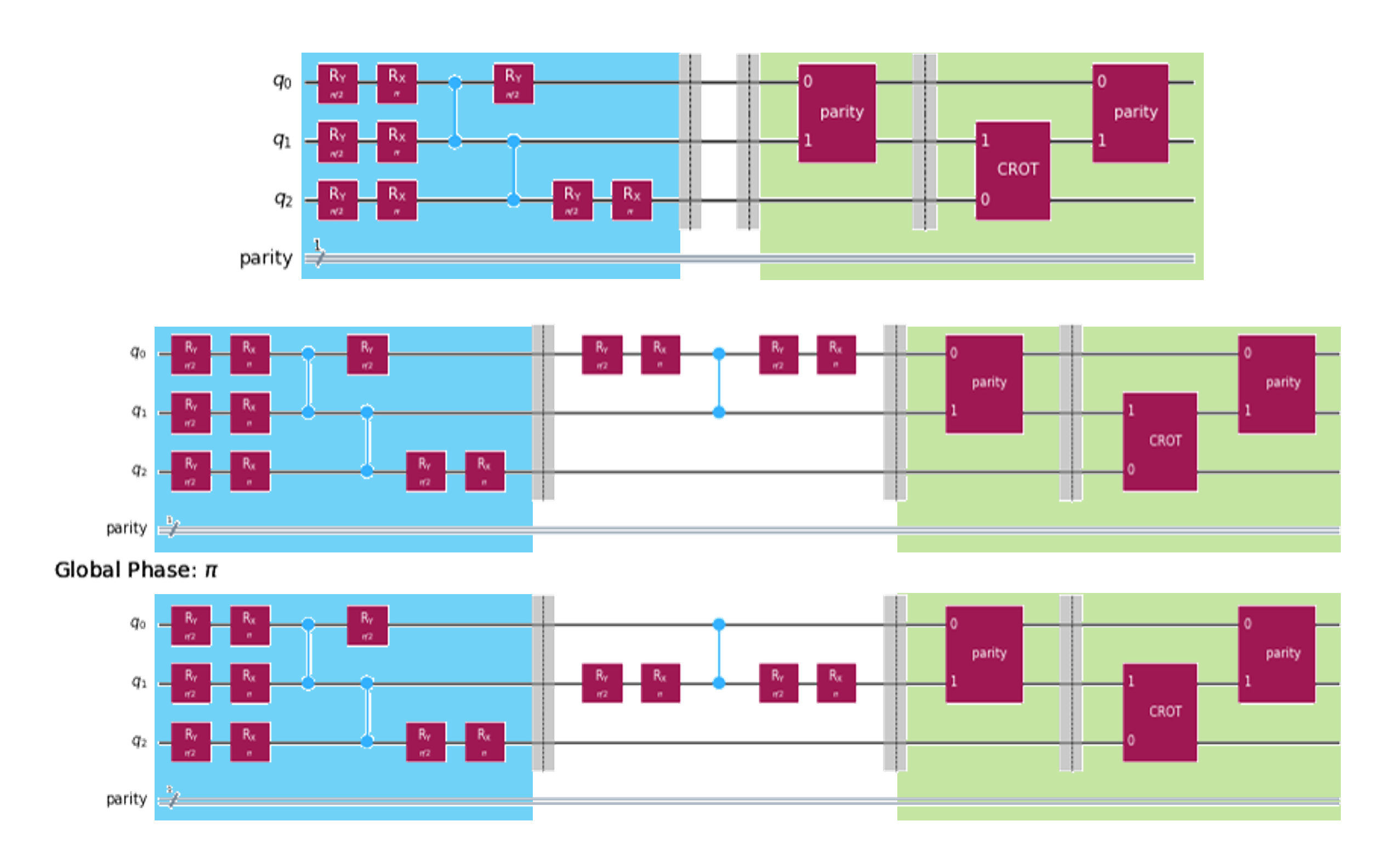}
			\caption{Tomography circuits}
			\label{fig:3qubitghzcircuit_b}
		\end{subfigure}
		\centering
		\caption[3-qubit GHZ circuits]{3-qubit GHZ state circuit and the corresponding reconstruction circuits. The readout of the inner qubit (the third qubit) is via a controlled rotation. Each circuit consists of the state preparation (blue background), reconstruction gates (white) and measurement (green).}
		\label{fig:3qubitghzcircuit}
	\end{figure}

\begin{figure}
		\centering
		\includegraphics[width=0.9\linewidth]{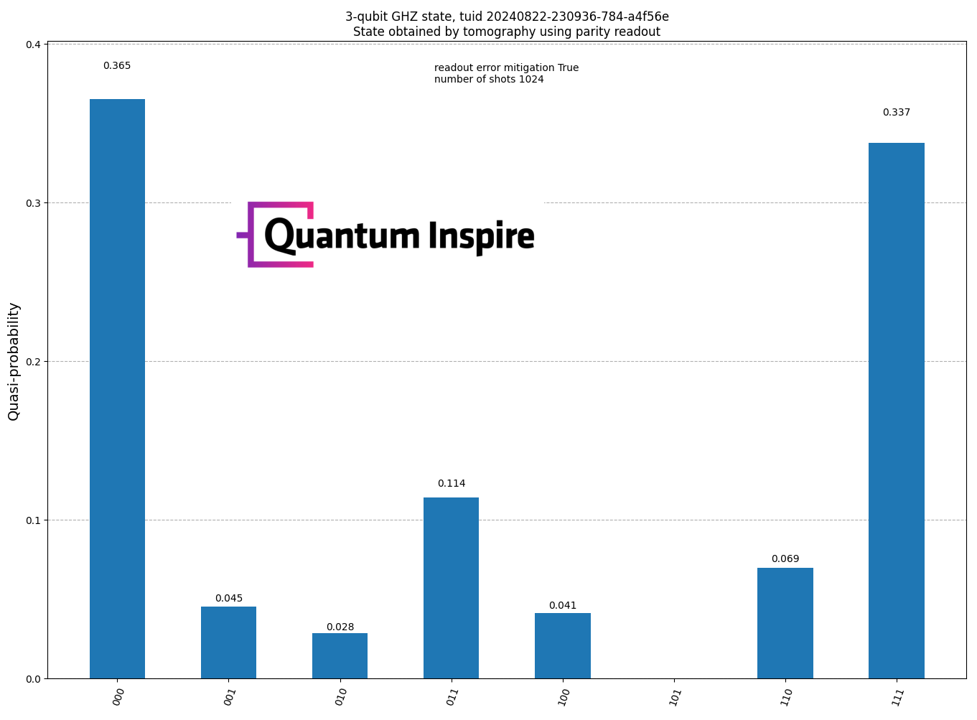}
		\caption{Reconstruction of a 3-qubit GHZ state on the Quantum Inspire Spin-X device.}
		\label{fig:3qubitghzusingparitytomography}
	\end{figure}

	\rsection{Device results}
	We have implemented the reconstruction method on a six-quantum dot spin qubit device. Native gates on this device are the $R_x$ and $R_y$ gates, the $R_z$ gate (virtual, hence noiseless) and the $CZ$ gate. We created a 3-qubit GHZ state using the circuit diagram shown in \figureref{fig:3qubitghzcircuit_a}. The circuit is transpiled into circuit with native gates only (\figureref{fig:3qubitghzcircuit_b}). The resulting reconstruction for the GHZ state is shown in \figureref{fig:3qubitghzusingparitytomography}. The reconstruction has trace norm 0.298, which is higher than the typical trace norm expected from the simulation. This can be explained by the native 2-qubit gate having a low gate fidelity~\cite[section 9.3]{Nielsen2000} of only 80\% during the data acquisition.
	
\begin{figure}[ph]
		\centering
		\includegraphics[width=0.9\linewidth]{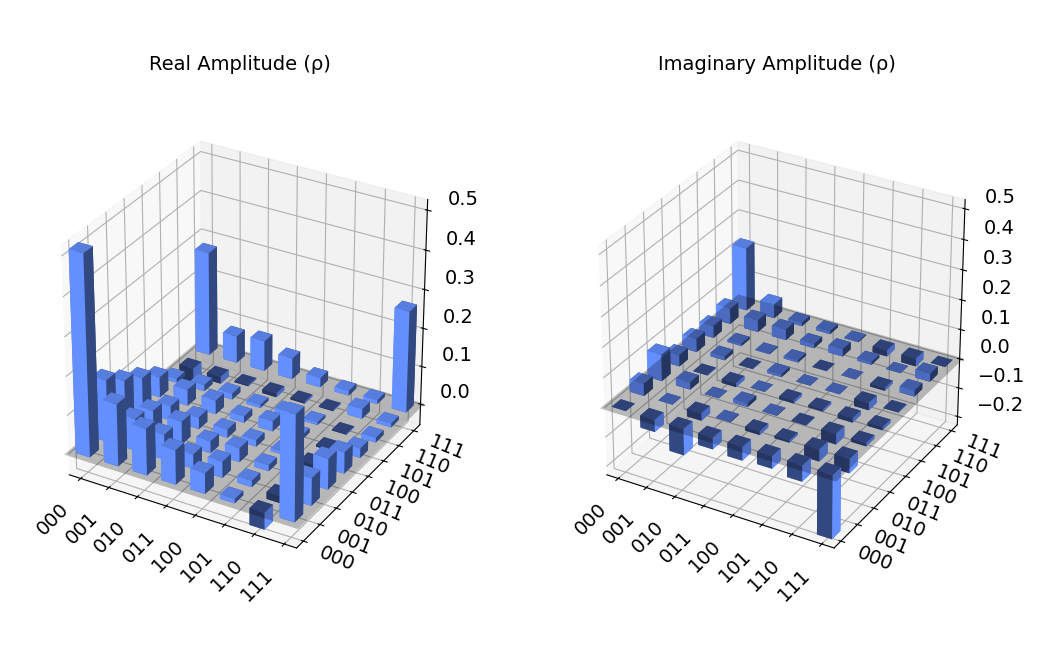}
		\caption[3-qubit GHZ state]{Quantum state tomography on a 3-qubit GHZ state using parity reconstruction. The density matrix fidelity was 0.661.
		Shown are the real (left) and imaginary (right) part of the density matrix.}
		\label{fig:3qubitghzqst20250114}
	\end{figure}

	We have also performed quantum state tomography on the 3-qubit GHZ state. This is a double reconstruction: we measure $27\cdot 3$ circuits and reconstruct from those the 27 results required for 3-qubit state tomography. We then perform state tomography to obtain the density matrix of the state shown in \figureref{fig:3qubitghzqst20250114}.
	

	\section{Discussion}
	
	We have shown that reconstruction using parity readout is a viable alternative to ancilla based readout. This result is relevant for spin-qubit systems where direct readout is often not available and for maximizing the number of computational qubits with a small computational overhead.
	Our simulations and device measurements further show that the overhead of tomography using parity readout is comparable to indirect readout using state transfer.
	
	Finally, spin states that are preserved during the parity measurements allow for an alternative reconstruction scheme. For spin quantum processors using Pauli spin blockade for readout, the odd parity states relax to the singlet state, whereas the even parity states $|00\rangle$ and $|11\rangle$ are preserved. This preservation enables a two-circuit readout scheme (with additional parity measurements and operations) to reconstruct the measurements~\cite{amitonov2024}.

	\section{Acknowledgements}
	
	This work was conducted with financial support from the Dutch National Growth Fund, which is part of the European Union’s NextGenerationEU recovery fund, and supported by the European Union’s Horizon 2020 research and innovation programme under the Grant Agreement No.951852 (QLSI project).

	\clearpage 
	\bibliographystyle{unsrt}
	\bibliography{references}

	\appendix

	\section{Device}
	
\begin{figure}
		\centering
		\includegraphics[width=0.87\linewidth]{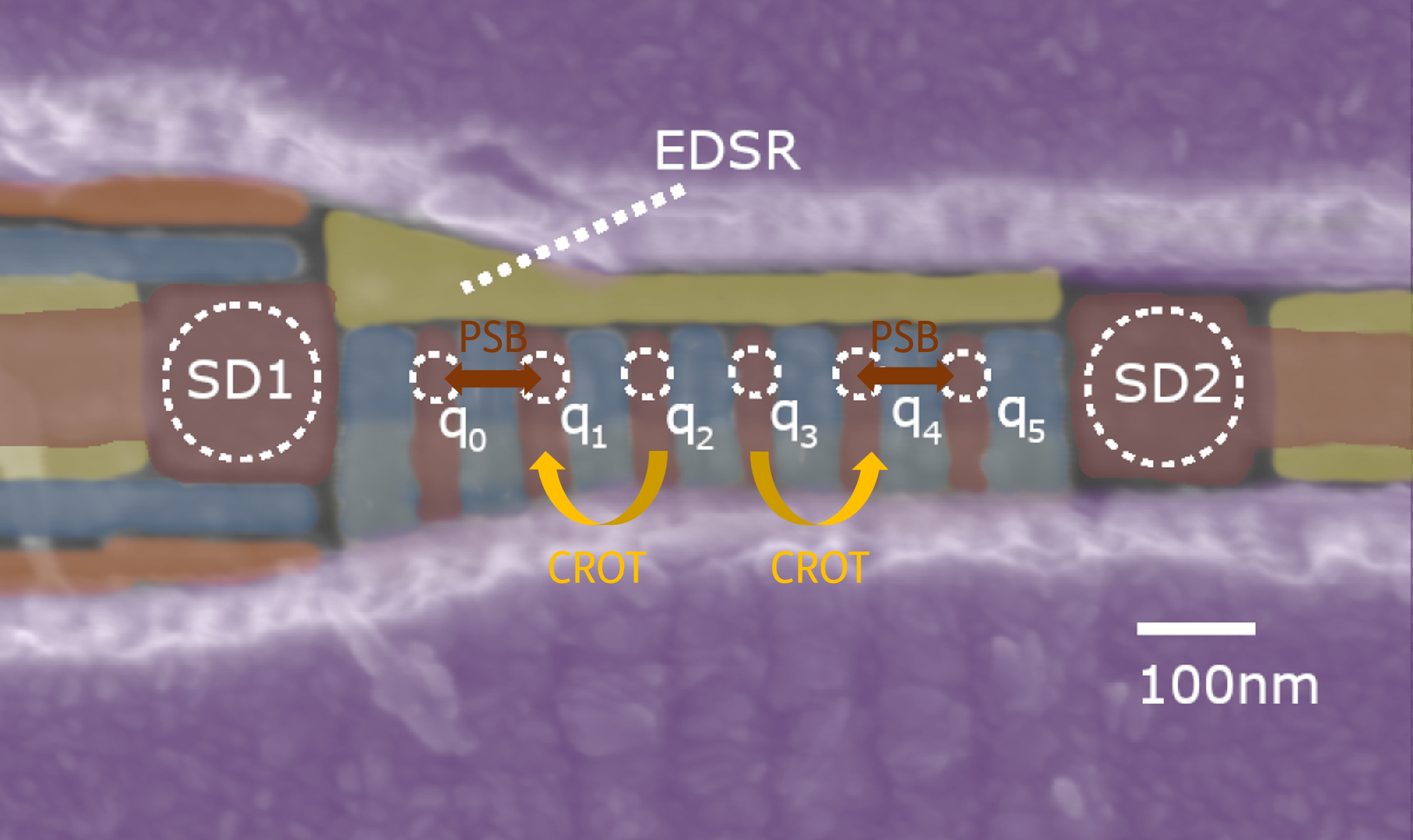}
		\caption{False color SEM image of the device nominally identical to the 6D2S device measured in this paper. q$_n$ indicate the quantum dots with few electron spins hosting the qubits. SD1 and SD2 are the sensing dots enabling the readout. Spin-to-charge conversion is done by Pauli spin blockade (PSB) using q$_1$, q$_2$ pair, and similarly using q$_4$, q$_5$ pair. Readout of the inner spins are enabled by CROT logic gates. Qubits are driven by electric-dipole spin resonance (EDSR) enabled by the micromagnet shown in purple creating a magnetic field gradient in vertical direction. The RF signal driving the qubits is delivered by the electrode in yellow indicated with EDSR. A similar device is reported in~\cite{Philips2022}.}
		\label{fig:6d2sdevicelayoutannotated}
	\end{figure}

	The device we use is a Si/SiGe heterostructure similar to the one in~\cite{Philips2022} (see~\figureref{fig:6d2sdevicelayoutannotated}).

	\section{Details on regularization}
	\label{appendix:regularization}
	We use regularization to avoid numerical instabilities around probability 0 and 1.
	The probability cost method we use is
	\begin{align}
		P(p) &= 
		\begin{cases}
			|p|^2   &  \text{ if $p < 0$} \\
			0   &  \text{ if $0 \leq p \leq 1$} \\
			(p-1)^2   &  \text{ if $p > 1$} \\
		\end{cases} \, .
	\end{align}
	This ensures any probability estimates outside the valid $[0, 1]$ region have a penalty term. For the regularization of the variance we use the regularized probability $r(p, \epsilon)$. The regularized probability is equal to the probability in the inner region, but regularized using a second order polynomial near 0 and 1. The regularized probability is defined as:
	\begin{align}
		r(p, \epsilon) &= 
		\begin{cases}
			\epsilon/2 & \quad \text{if $p<0$} \\
			\epsilon/2 + p^2/(2\epsilon) & \quad \text{if $0 \leq p \leq \epsilon$} \\
			p & \quad \text{if $\epsilon \leq p \leq 1-\epsilon $}  \\
			1- (\epsilon/2 + (1-p)^2/(2\epsilon)) & \quad \text{if $1-\epsilon \leq p \leq 1$} \\
			1-\epsilon/2 & \quad \text{if $p>1$} 
		\end{cases}  .
		\label{regularized probability} 
	\end{align}
	\begin{figure}
		\centering
		\includegraphics[width=0.82\linewidth]{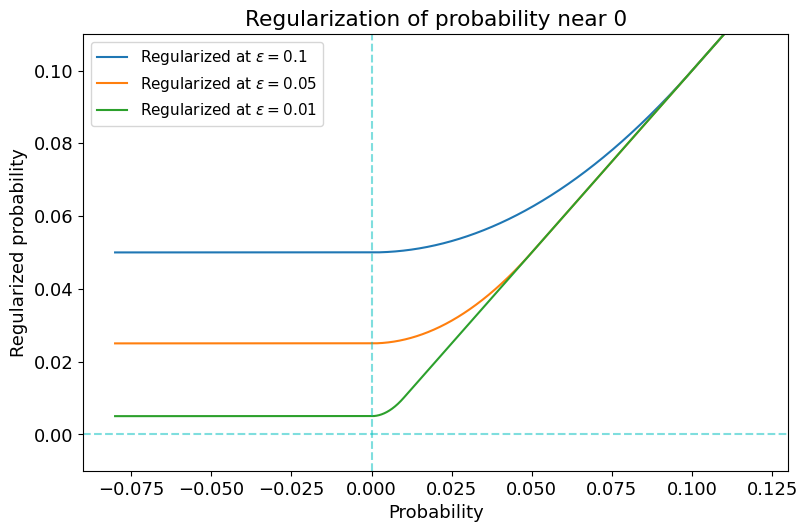}
		\caption{Probability regularization using a second order polynomial near 0 and 1.}
		\label{fig:regularization}
	\end{figure}
	The variance estimate is regularized as
	\begin{align}
		\sigma^2 &= r(p, \epsilon) (1-r(p, \epsilon))/ N \, ,
	\end{align}
	with $N$ the number of shots. The regularization ensures that the variance estimate is non-zero for all estimates $p$.

	\section{Simulation details}
	\label{appendix:simulation details}
	The Spin-X simulations have been performed using Qiskit with noise added to the single-qubit gates, two-qubit gates and the readout operation.
	We compile every circuits to native gates $\Rx$, $\Ry$, $\Rz$ and CZ. We add depolarization
	noise~\cite[section 9.4.3]{Nielsen2000} with parameter $p=0.05$ to all single-qubit rotations and in addition add an over-rotation of 1 degree to the first qubit. For 2-qubit CZ gate we add depolarizing noise on each of the qubits involved with parameter $p=0.1$.
	For the readout operation, we simulate readout with fidelity 0.97.
	To simulate the parity readout we replace the parity readout operation on a qubit pair $(a,b)$ by the combination of a noiseless $\mathrm{CNOT}(a,b)$ and a direct readout on qubit $b$.

\section{Selection of reconstruction circuits}
\label{appendix:reconstruction circuits}
\label{appendix:dof}

A lower bound on the minimal number of circuits required for reconstruction of either a z-basis measurement or density matrix can be obtained by analyzing the degrees of freedom. The z-basis measurement has $2^\nq-1$ degrees of freedom, the density matrix has $2^{2\nq}-2$ real degrees of freedom. This is summarized in~\tableref{table:dof}.
		\renewcommand{\arraystretch}{1.2}
\begin{table}
	\begin{tabular}{|c|c|c|c|c|}
		\hline
		& \multicolumn{4}{c|}{\bfseries{Number of qubits}}  \\
		\hline 
		& 1 & 2 & 3 & 6 \\
		\hline
		Dimension of state vector & 2 & 4 & 8 & 64 \\
		\hline
		Real degrees of freedom in the state vector & 2 & 6 & 14 & 126 \\
		\hline
		Density matrix shape & $2\times 2$ & $4\times 4$ & $8\times 8$ & $64\times 64$ \\
		\hline
		DM real degrees of freedom & 3  & 15  & 63 & 4095 \\
		\hline
		Z-basis measurement real degrees of freedom & 1  & 3  & 7 & 63 \\
		\hline
		\hline
		Minimal number of circuits for z-basis & -  & 3  & 3 & 3 \\
		\hline
	Minimal number of circuits for DM & -  & 3  & 21 & 133 \\
		\hline
	\end{tabular}
	\caption{State vector and density matrix dimensions. Number of circuits required for reconstruction.}
	\label{table:dof}
\end{table}

The reconstruction circuits correspond to observables to be measured. Note that every observable can be expressed as a sum of Pauli terms (see~\cite{enwiki:pauli_matrices}), e.g. as a sum of
terms from $\{\pI,\pX,\pY,\pZ\}^{\otimes k}$.
But any term involving a Pauli $\pY$ provides no information about the Z-basis measurements, since
if we have real coefficients, expectation of the Pauli $\pY$ is 0. Terms involving terms \pX suffer from the issue that their expectation value is non-linear in the state vectors.
E.g.~$E_{\psi+\phi}(X) \neq E_\psi(X) + E_\phi(X)$ (this is in contrast with the density matrices where
we have $E_{\rho+\rho'}(\pX) = \tr( (\rho+\rho') X) = \tr( \rho \pX) +\tr( \rho' \pX) = E_\rho(X) + E_{\rho'} (X) $).
This prevents our linear construction to work properly.
So we reduce (without loss of generality) our set of observables to a sum of terms $\pI \otimes \pI$,  $\pI \otimes \pZ$,  $\pZ \otimes \pI$,  and $\pZ \otimes \pZ$.
These observables have the property that the expectation of the sum of two states equals the sum of the expectations of the individual states.
Since $\pI \otimes \pI$ is constant and we need at least 3 independent circuits our choice of circuits up to equivalence is unique. In other words, from a quantum information point of view there is only one choice in circuits. From an implementation point of view there are some differences, as some observables require more gates to implement on a given backend. In the main text a choice of observables suitable for typical hardware is selected.

For a system with 3 qubits for which the first two have parity readout and the last one direct readout, we can use the following inversion matrix

	\begin{align}
	M &= \begin{pmatrix}
		1 & 0 & 0 & 1 & 0 & 0 & 0 & 0  \\
		0 & 0 & 0 & 0 & 1 & 0 & 0 & 1  \\
		0 & 1 & 1 & 0 & 0 & 0 & 0 & 0  \\
		0 & 0 & 0 & 0 & 0 & 1 & 1 & 0  \\
		1 & 0 & 1 & 0 & 0 & 0 & 0 & 0  \\
		0 & 0 & 0 & 0 & 1 & 0 & 1 & 0  \\
		0 & 1 & 0 & 1 & 0 & 0 & 0 & 0  \\
		0 & 0 & 0 & 0 & 0 & 1 & 0 & 1  \\
		1 & 1 & 0 & 0 & 0 & 0 & 0 & 0  \\
		0 & 0 & 0 & 0 & 1 & 1 & 0 & 0  \\
		0 & 0 & 1 & 1 & 0 & 0 & 0 & 0  \\
		0 & 0 & 0 & 0 & 0 & 0 & 1 & 1  
	\end{pmatrix}
	\label{equation:M3} .
\end{align}

	\section{Gate set tomography}
	\label{appendix:gst}
	The gate set tomography experiments have been performed with PyGSTi~\cite{pyGSTi}. We use the \code{smq2Q\_XYCPHASE} modelpack and germs up to length 8. Our experiments use a simulator with noise on the readout and the single- and two-qubit gates. We performed 3 reconstructions: direct readout, direct readout with readout error mitigation~\cite{Maciejewski2020} and finally our reconstruction method.
	All three methods successfully reconstruct the single- and two-qubit gates and their fidelities. This is to be expected, as all methods only differ in the readout (the initialization and gate operations are identical). 
	Both the readout error mitigation and the reconstruction method show a large model violation. This is an indication that the readout part in these methods does not satisfy the pure quantum measurement operator statistics. Usually a high model violation indicates a non-Markovian system, but that is not the case in our simulation.
	Finally the SPAM error is the most interesting. Our baseline has estimated SPAM error 0.964. The readout error mitigation has estimated SPAM probability 0.995. So the readout error mitigation created a better SPAM, at the cost of introducing statistics that does not match the pure quantum measurement operator statistics (hence the high model violation). Our reconstruction method has only a slightly lower SPAM probability at 0.957.

	\section{State and probability metrics}
	\label{appendix:metrics}
	\makeatletter
	\newcommand{\owntag}[2][\relax]{
		\ifx#1\relax\relax\def\owntag@name{#2}\else\def\owntag@name{#1}\fi
		\refstepcounter{equation}\tag{\theequation, #2}%
		\expandafter\ltx@label\expandafter{eq:\owntag@name}%
		\edef\@currentlabel{\theequation, #2}\expandafter\ltx@label\expandafter{Eq:\owntag@name}%
		\def\@currentlabel{#2}\expandafter\ltx@label\expandafter{tag:\owntag@name}%
	}
	\makeatother
	Two natural metrics for probability distributions, quantum states or density matrices are the trace norm and state fidelity.
	For probability distributions we define
	\begin{align}
		\Dtrace(p, q) &= \sum |p_i - q_i| / 2  \refstepcounter{equation} \tag{trace distance~\cite[equation 9.1]{Nielsen2000}, \theequation} \\
		\F(p, q) &= \sum \sqrt{p_i q_i}  \refstepcounter{equation}\tag{fidelity~\cite[equation 9.2]{Nielsen2000},\theequation}
	\end{align}
	For pure states the state fidelity is equivalent to the trace fidelity. See~\cite[section 9.2.3]{Nielsen2000}.
	For two states $|\psi\rangle = \sum_i a_i |i\rangle$,  $|\phi\rangle = \sum_i b_i |i\rangle$
	the state fidelity is defined as
	\begin{align}
		\F(|\psi\rangle, |\phi\rangle) &= |\langle \psi | \phi \rangle|^2 = |\sum_i {a_i^\dagger b_i} |^2 
	\end{align}
	With this definition states $|0\rangle + |1\rangle$ and $|0\rangle - |1\rangle$ have fidelity zero. But when performing a Z-basis measurement
	these states have the same probability distribution for the measurement outcomes. So from a reconstruction point of view,  
	$|0\rangle + |1\rangle$ and $|0\rangle - |1\rangle$ are equally good reconstructions. For that reason we are looking at the trace norm of the 
	probability distributions $|a_i| ^2$.
	
	For pure states $|a\rangle$, $|b\rangle$ the trace distance and fidelity are related by
	\begin{align}
		\Dtrace(|a\rangle, |b\rangle) = \sqrt{1-F(|a\rangle, |b\rangle)^2} \, .
	\end{align}
	So the choice between trace distance and fidelity is a matter of scale. Also see~\cite{wiki:fidelity_of_states}.

\end{document}